# Sparse Codes for Speech Predict Spectrotemporal Receptive Fields in the Inferior Colliculus

Nicole L. Carlson[1,2], Vivienne L. Ming[1], Michael Robert DeWeese[1,2,3]*

1 Redwood Center for Theoretical Neuroscience, University of California, Berkeley, California, United States of America, 2 Department of Physics, University of California, Berkeley, California, United States of America, 3 Helen Wills Neuroscience Institute, University of California, Berkeley, California, United States of America

## Abstract

We have developed a sparse mathematical representation of speech that minimizes the number of active model neurons needed to represent typical speech sounds. The model learns several well-known acoustic features of speech such as harmonic stacks, formants, onsets and terminations, but we also find more exotic structures in the spectrogram representation of sound such as localized checkerboard patterns and frequency-modulated excitatory subregions flanked by suppressive sidebands. Moreover, several of these novel features resemble neuronal receptive fields reported in the Inferior Colliculus (IC), as well as auditory thalamus and cortex, and our model neurons exhibit the same tradeoff in spectrotemporal resolution as has been observed in IC. To our knowledge, this is the first demonstration that receptive fields of neurons in the ascending mammalian auditory pathway beyond the auditory nerve can be predicted based on coding principles and the statistical properties of recorded sounds.

Citation: Carlson NL, Ming VL, DeWeese MR (2012) Sparse Codes for Speech Predict Spectrotemporal Receptive Fields in the Inferior Colliculus. PLoS Comput Biol 8(7): e1002594. doi:10.1371/journal.pcbi.1002594

Editor: Tim Behrens, University of Oxford, United Kingdom

Received November 16, 2011; Accepted May 18, 2012; Published July 12, 2012

Copyright: © 2012 Carlson et al. This is an open-access article distributed under the terms of the Creative Commons Attribution License, which permits unrestricted use, distribution, and reproduction in any medium, provided the original author and source are credited.

Funding: NLC was supported by a National Science Foundation graduate fellowship; MRD gratefully acknowledges support from the National Science Foundation (www.nsf.gov), the McKnight Foundation (www.mcknight.org), the McDonnell Foundation (www.jsmf.org), and the Hellman Family Faculty Fund. The funders had no role in study design, data collection and analysis, decision to publish, or preparation of the manuscript.

Competing Interests: The authors have declared that no competing interests exist.

* E-mail: deweese@berkeley.edu

## Introduction

Our remarkable ability to interpret the highly structured sounds in our everyday environment suggests that auditory processing in the brain is somehow specialized for natural sounds. Many authors have postulated that the brain tries to transmit and encode information efficiently, so as to minimize the energy expended [1], reduce redundancy [2–4], maximize information flow [5–8], or facilitate computations at later stages of processing [9], among other possible objectives. One way to create an efficient code is to enforce population sparseness, having only a few active neurons at a time. Sparse coding schemes pick out those features of a signal — those features that occur much more often than chance — which can then be used to efficiently represent a complex signal with few active neurons.

The principle of sparse coding has led to important insights into the neural encoding of visual scenes within the primary visual cortex (V1). Sparse coding of natural images revealed local, oriented edge-detectors that qualitatively match the receptive fields of simple cells in V1 [10]. More recently, overcomplete sparse coding schemes have uncovered a greater diversity of features that more closely matches the full range of simple cell receptive field shapes found in V1 [11]. An encoding is called overcomplete if the number of neurons available to represent the stimulus is larger than the dimensionality of the input. This is a biologically realistic property for a model of sensory processing because information is encoded by increasing numbers of neurons as it travels from the optic nerve to higher stages in the visual pathway, just as auditory sensory information is encoded by increasing numbers of neurons as it travels from the auditory nerve to higher processing stages [12].

Despite experimental evidence for sparse coding in the auditory system [13,14], there have been fewer theoretical sparse coding studies in audition than in vision. However, there has been progress, particularly for the earliest stages of auditory processing. Sparse coding of raw sound pressure level waveforms of natural sounds produced a "dictionary" of acoustic filters closely resembling the impulse response functions of auditory nerve fibers [15,16]. Acoustic features learned by this model were best fit to the neural data for a particular combination of animal vocalizations and two subclasses of environmental sounds. Intriguingly, they found that training on speech alone produced features that were just as well-fit to the neural data as the optimal combination of natural sounds, suggesting that speech provides the right mixture of acoustic features for probing and predicting the properties of the mammalian auditory system.

Another pioneering sparse coding study [17] took as its starting point speech that was first preprocessed using a model of the cochlea — one of several so-called cochleagram representations of sound. This group found relatively simple acoustic features that were fairly localized in time and frequency as well as some temporally localized harmonic stacks. These results were roughly consistent with some properties of receptive fields in primary auditory cortex (A1), but modeled responses did not capture the majority of the specific shapes of neuronal spectrotemporal receptive fields (STRFs; [18]) reported in the literature. That study only considered undercomplete dictionaries, and it focused solely on a "soft" sparse coding model that minimized the mean

## Author Summary

The receptive field of a neuron can be thought of as the stimulus that most strongly causes it to be active. Scientists have long been interested in discovering the underlying principles that determine the structure of receptive fields of cells in the auditory pathway to better understand how our brains process sound. One possible way of predicting these receptive fields is by using a theoretical model such as a sparse coding model. In such a model, each sound is represented by the smallest possible number of active model neurons chosen from a much larger group. A primary question addressed in this study is whether the receptive fields of model neurons optimized for natural sounds will predict receptive fields of actual neurons. Here, we use a sparse coding model on speech data. We find that our model neurons do predict receptive fields of auditory neurons, specifically in the Inferior Colliculus (midbrain) as well as the thalamus and cortex. To our knowledge, this is the first time any theoretical model has been able to predict so many of the diverse receptive fields of the various cell-types in those areas.

activity of the model's neurons, as opposed to "hard" sparse models that minimize the number of active neurons.

The same group also considered undercomplete, soft sparse coding of spectrograms of speech [19], which did yield some STRFs showing multiple subfields and temporally modulated harmonic stacks, but the range of STRF shapes they reported was still modest compared with what has been seen experimentally in auditory midbrain, thalamus, or cortex. Another recent study considered sparse coding of music [20] in order to develop automated genre classifiers.

To our knowledge, there are no published studies of complete or overcomplete, sparse coding of either spectrograms or cochleograms of speech or natural sounds. We note that one preliminary sparse coding study utilizing a complete dictionary trained on spectrograms did find STRFs resembling formants, onset-sensitive neurons, and harmonic stacks (J. Wang, B.A. Olshausen, and V.L. Ming, COSYNE 2008) but they did not obtain novel acoustic features, nor any that closely resembled STRFs from the auditory system.

Our goal is two-fold. First, we test whether an overcomplete, hard sparse coding model trained on spectrograms of speech can more fully reveal the structure of natural sounds than previous models. Second, we ask whether our model can accurately predict receptive fields in the ascending auditory pathway beyond the auditory nerve. We have found that, when trained on spectrograms of human speech, an overcomplete, hard sparse coding model does learn features resembling those of STRF shapes previously reported in the inferior colliculus (IC), as well as auditory thalamus and cortex. Moreover, our model exhibits a similar tradeoff in spectrotemporal resolution as previously reported in IC. Finally, our model has identified novel acoustic features for probing the response properties of neurons in the auditory pathway that have thus far resisted classification and meaningful analysis.

## Results

### Sparse Coding Model of Speech

In order to uncover important acoustic features that can inform us about how the nervous system processes natural sounds, we have developed a sparse coding model of human speech (see Methods for details). As illustrated in **Fig. 1**, raw sound pressure level waveforms of recorded speech were first preprocessed by one of two simple models of the peripheral auditory system. The first of these preprocessing models was the spectrogram, which can be thought of as the power spectrum of short segments of the original waveform at each moment in time. We also explored an alternative preprocessing step that was meant to more accurately model the cochlea [21,22]; the original waveform was sent through a filter bank with center frequencies based on the properties of cochlear nerve fibers. Both models produced representations of the waveform as power at different frequencies over time. The spectrograms (cochleograms) were then separated into segments of length 216 ms (250 ms). Because of the high dimensionality of these training examples, we performed principal components analysis (PCA) and retained only the first two hundred components to reduce the dimensionality (from $256 \times 25 = 6,400$ values down to 200), as was done previously in some visual [23] and auditory [17] sparse coding studies; the latter group also performed the control of repeating their analysis without the PCA step and they found that their results did not change.

We then trained a "dictionary" of model neurons that could encode this data using the Locally Competitive Algorithm (LCA), a recently developed sparse encoding algorithm [24]. This flexible algorithm allowed us to approximately enforce either the so-called "hard" sparseness (L0 sparseness; minimizing the number of simultaneously active model neurons) or "soft" sparseness (L1 sparseness; minimizing the sum of all simultaneous activity across all model neurons) during encoding by our choice of thresholding function. Additionally, we explored the effect of dictionary over-completeness (with respect to the number of principal components) by training dictionaries that were half-complete, complete, or overcomplete (two or four times). Following training, the various resulting dictionaries were analyzed for cell-types and compared to experimental receptive fields reported in the literature.

### Cochleogram-Trained Models

In general, training our network on cochleogram representations of speech resulted in smooth and simple shapes for the learned receptive fields of model neurons. Klein and colleagues [17] used a sparse coding algorithm that imposed an L1-like sparseness constraint to learn a half-complete dictionary of cochleograms. Their dictionary elements consisted of harmonic stacks at the lower frequencies and localized elements at the higher frequencies. To make contact with these results, we trained a half-complete L0-sparse dictionary on cochleograms and compared the response properties of our model neurons with those of the previous study. The resulting dictionary (**Fig. 2**) consists of similar shapes to this previous work with the exception of one "onset element" in the upper left (this is the least used of all of the elements from this dictionary). Subsequent simulations revealed that the form of the dictionary is strongly dependent on the degree of overcompleteness. Even a complete dictionary exhibits a greater diversity of shapes than this half-complete dictionary (**Fig. S11**). This was true for L1-sparse dictionaries trained with LCA [24] or with Sparsenet [10] (**Figs. S15** and **S19**).

The inability of the half-complete dictionary to produce the more complex receptive field shapes of the complete dictionary, such as those resembling STRFs measured in IC, or those in auditory thalamus or cortex, suggests that overcompleteness in those regions is crucial to the flexibility of their auditory codes.

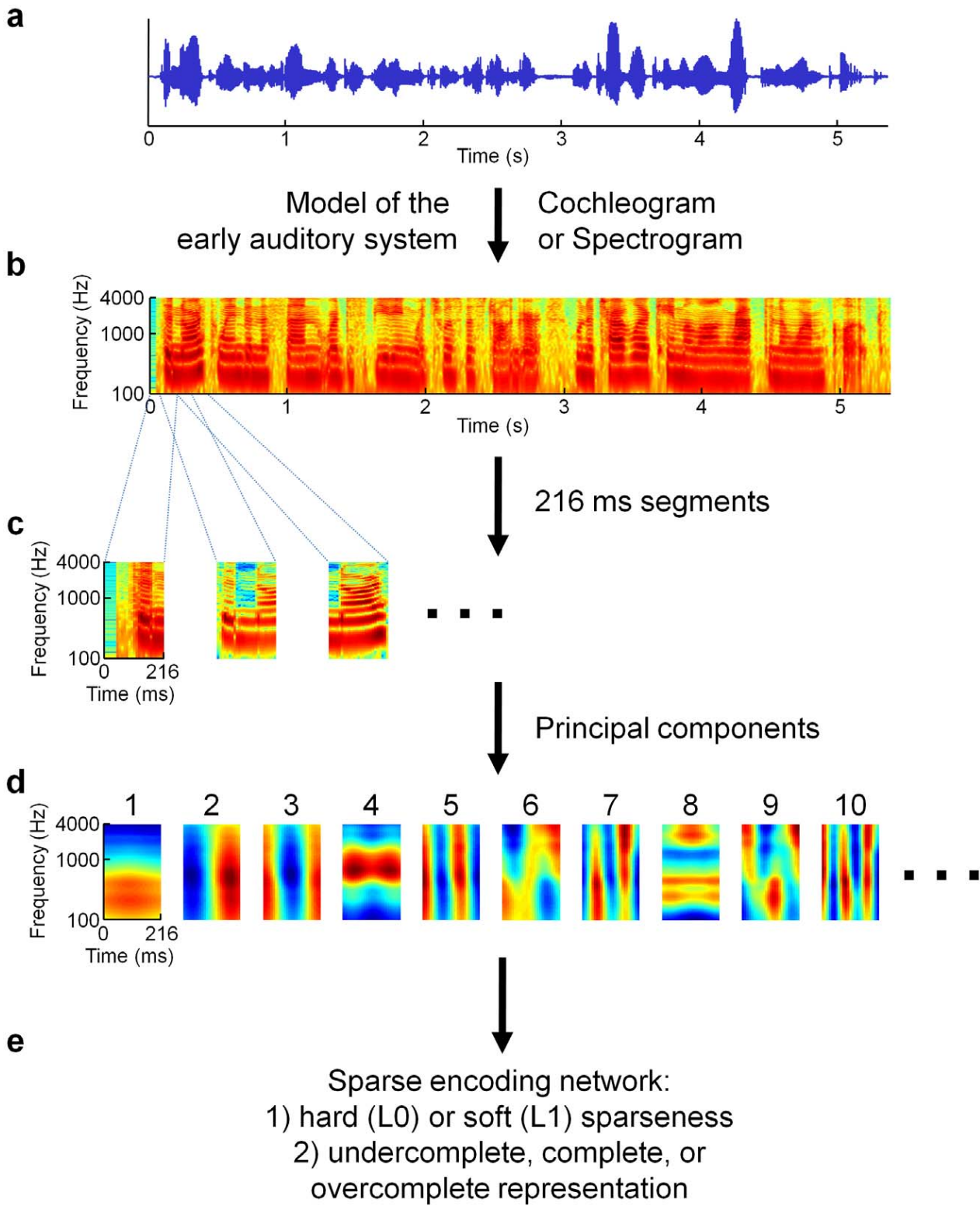

**Figure 1. Schematic illustration of our sparse coding model.** (**a**) Stimuli used to train the model consisted of examples of recorded speech. The blue curve represents the raw sound pressure waveform of a woman saying, "The north wind and the sun were disputing which was the stronger, when a traveler came along wrapped in a warm cloak." (**b**) The raw waveforms were first put through one of two preprocessing steps meant to model the earliest stages of auditory processing to produce either a spectrogram or a "cochleogram" (not shown; see Methods for details). In either case, the power spectrum across acoustic frequencies is displayed as a function of time, with warmer colors indicating high power content and cooler colors indicating low power. (**c**) The spectrograms were then divided into overlapping 216 ms segments. (**d**) Subsequently, principal components analysis (PCA) was used to project each segment onto the space of the first two hundred principal components (first ten shown), in order to reduce the dimensionality of the data to make it tractable for further analysis while retaining its basic structure [17]. (**e**) These projections

were then input to a sparse coding network in order to learn a "dictionary" of basis elements analogous to neuronal receptive fields, which can then be used to form a representation of any given stimulus (i.e., to perform inference). We explored networks capable of learning either "hard" (L0) sparse dictionaries or "soft" (L1) sparse dictionaries (described in the text and Methods) that were undercomplete (fewer dictionary elements than PCA components), complete (equal number of dictionary elements), or over-complete (greater number of dictionary elements).
doi:10.1371/journal.pcbi.1002594.g001

## Spectrogram-Trained Models

The spectrogram-trained dictionaries provide a much richer and more diverse set of dictionary elements than those trained on cochleograms. We display representative elements of the different categories of shapes found in a half-complete L0-sparse spectrogram dictionary (**Fig. 3a–f**) along with a histogram of the usage of the elements (**Fig. 3g**) when used to represent individual sounds drawn from the training set (i.e., during inference). Interestingly, we find that the different qualitative types of neurons separate according to their usage into a series of rises and plateaus. The least used elements are the harmonic stacks (**Fig. 3a**), which is perhaps unsurprising since, in principle, only one of them needs to be active at many points in time for a typical epoch of a recording from a single human speaker. We note that, while such harmonic stack receptive fields are apparently rare in the colliculus, thalamus, and cortex, they are well represented in the dorsal cochlear nucleus (DCN) (e.g., see Fig. 5b in [25]). The neighboring flat region consists of onset elements (**Fig. 3b**), which contain broad frequency subfields that change abruptly at one moment in time. These neurons were all used approximately equally often across the training set since it is equally probable that a stimulus will initiate transient will occur any time during the 216 ms time window.

The third region consists of more complex harmonic stacks that contain low-power subfields on the sides (**Fig. 3c**), a feature sometimes referred to as "temporal inhibition" or "band-passed inhibition" when observed in neural receptive fields; we will refer to this as "suppression" rather than inhibition to indicate that the model is agnostic as to whether these suppressed regions reflect direct synaptic inhibition to the neuron, rather than a decrease in excitatory synaptic input. The next flat region represents stimulus onsets, or ON-type cells, that tend to be more localized in frequency (**Fig. 3d**). The fifth group of elements is reminiscent of formants (**Fig. 3e**), which are resonances of the vocal tract that appear as characteristic frequency modulations common in speech. Formants are modulations "on top of" the underlying harmonic stack, often consisting of pairs of subfields that diverge or converge over time in a manner that is not consistent with a pair of harmonics rising or falling together due to fluctuations in the fundamental frequency of the speaker's voice. The final region consists of the most active neurons (**Fig. 3f**), which are highly localized in time and frequency and exhibit tight checkerboard-like patterns of excitatory and suppressive subfields (**Fig. 3f**). These features are exciting because they are similar to experimentally measured receptive field shapes that to our knowledge have not previously been theoretically predicted, as discussed below.

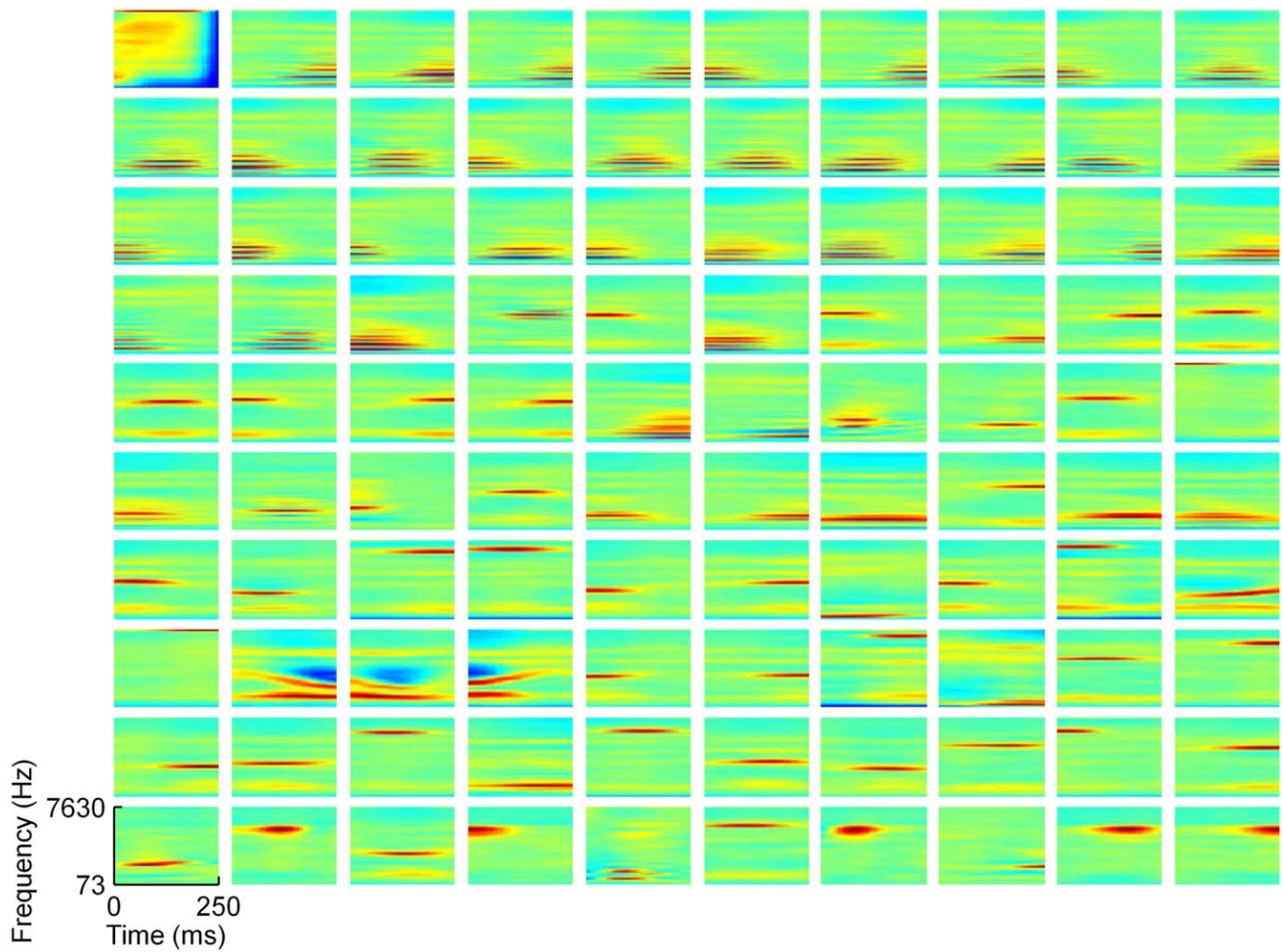

**Figure 2. A half-complete sparse coding dictionary trained on cochleogram representations of speech.** This dictionary exhibits a limited range of shapes. The full set of 100 elements from a half-complete, L0-sparse dictionary trained on cochleograms of human speech resemble those found in a previous study [17]. Nearly all elements are extremely smooth, with most consisting of a single frequency subfield or an unmodulated harmonic stack. Each rectangle can be thought of as representing the spectro-temporal receptive field (STRF) of a single element in the dictionary (see Methods for details); time is plotted along the horizontal axis (from 0 to 250 ms), and log frequency is plotted along the vertical axis, with frequencies ranging from 73 Hz to 7630 Hz. Color indicates the amount of power present at each frequency at each moment in time, with warm colors representing high power and cool colors representing low power. Each element has been normalized to have unit Euclidean length. Elements are arranged in order of their usage during inference (i.e., when used to represent individual sounds drawn from the training set) with usage increasing from left to right along each row, and all elements of lower rows used more than those of higher rows.
doi:10.1371/journal.pcbi.1002594.g002

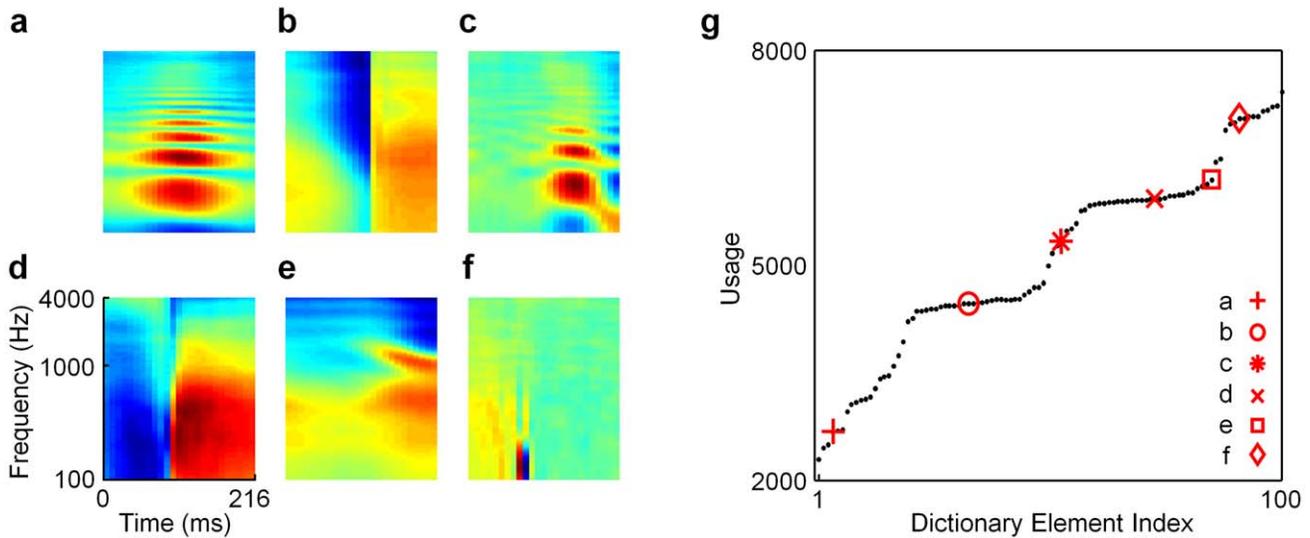

**Figure 3. A half-complete, L0-sparse dictionary trained on spectrograms of speech.** This dictionary exhibits a variety of distinct shapes that capture several classes of acoustic features present in speech and other natural sounds. (**a**–**f**) Selected elements from the dictionary that are representative of different types of receptive fields: (**a**) a harmonic stack; (**b**) an onset element; (**c**) a harmonic stack with flanking suppression; (**d**) a more localized onset/termination element; (**e**) a formant; (**f**) a tight checkerboard pattern (see **Fig. S1** for the full dictionary). Each rectangle represents the spectro-temporal receptive field (STRF) of a single element in the dictionary; time is plotted along the horizontal axis (from 0 to 216 msec) and log frequency is plotted along the vertical axis, with frequencies ranging from 100 Hz to 4000 Hz. (**g**) A graph of the usage of the dictionary elements showing that the different types of receptive field shapes separate based on usage into a series of rises and plateaus; red symbols indicate where each of the examples from panels **a**–**f** fall on the graph. The vertical axis represents the number of stimuli that required a given dictionary element in order to be represented accurately during inference.
doi:10.1371/journal.pcbi.1002594.g003

### Overcompleteness Affects Learned Features

Analogous to sparse coding studies in vision [11,26], we find that the degree of overcompleteness strongly influences the range and complexity of model STRF shapes.

**Fig. 4** presents representative examples of essentially all distinct cell types found in a four-times overcomplete L0-sparse dictionary trained on spectrograms. Features in the half-complete dictionary do appear as subsets of the larger dictionaries (**Fig. 4a, c, e, g, l**), but with increasing overcompleteness more complex features emerge, exhibiting richer patterns of excitatory and suppressive subfields. In general, optimized overcomplete representations can better capture structured data with fewer active elements, since the greater number of elements allows for important stimulus features to be explicitly represented by dedicated elements. In the limit of an infinite dictionary, for example, each element could be used as a so-called "grandmother cell" that perfectly represents a single, specific stimulus while all other elements are inactive.

Novel features that were not observed in smaller dictionaries include: an excitatory harmonic stack flanked by a suppressive harmonic stack (**Fig. 4b**); a neuron excited by low frequencies (**Fig. 4d**); a neuron sensitive to two middle frequencies (**Fig. 4f**); a localized but complex excitatory subregion followed by a suppressive subregion that is strongest for high frequencies (**Fig. 4h**); a checkerboard pattern with roughly eight distinct subregions (**Fig. 4i**); a highly temporally localized OFF-type neuron (**Fig. 4j**); and a broadband checkerboard pattern that extends for many cycles in time (**Fig. 4k**). Several of these features resemble STRFs reported in IC and further up the auditory pathway (see the "Predicting acoustic features that drive neurons in IC and later stages in the ascending auditory pathway" section below). One interesting property of the checkerboard units is that they are nearly separable in space and time [27], which has been studied for these and other types of neurons in ferret IC [28]. This is in contrast to some of the other model STRF shapes we have found, such as the example shown in Fig. 4e, which contains a strong diagonal subfield that is not well described by a product of independent functions of time and frequency.

As in the case of the half-complete dictionary (**Fig. 3**), the different classes of receptive field shapes segregate as a function of usage even as more intermediary shapes appear (see **Fig. S4** for the entire four-times overcomplete dictionary). However, the plateaus and rises evident in the usage plot for the half-complete dictionary (**Fig. 3g**) are far less distinct for the overcomplete representation (**Fig. 4m**).

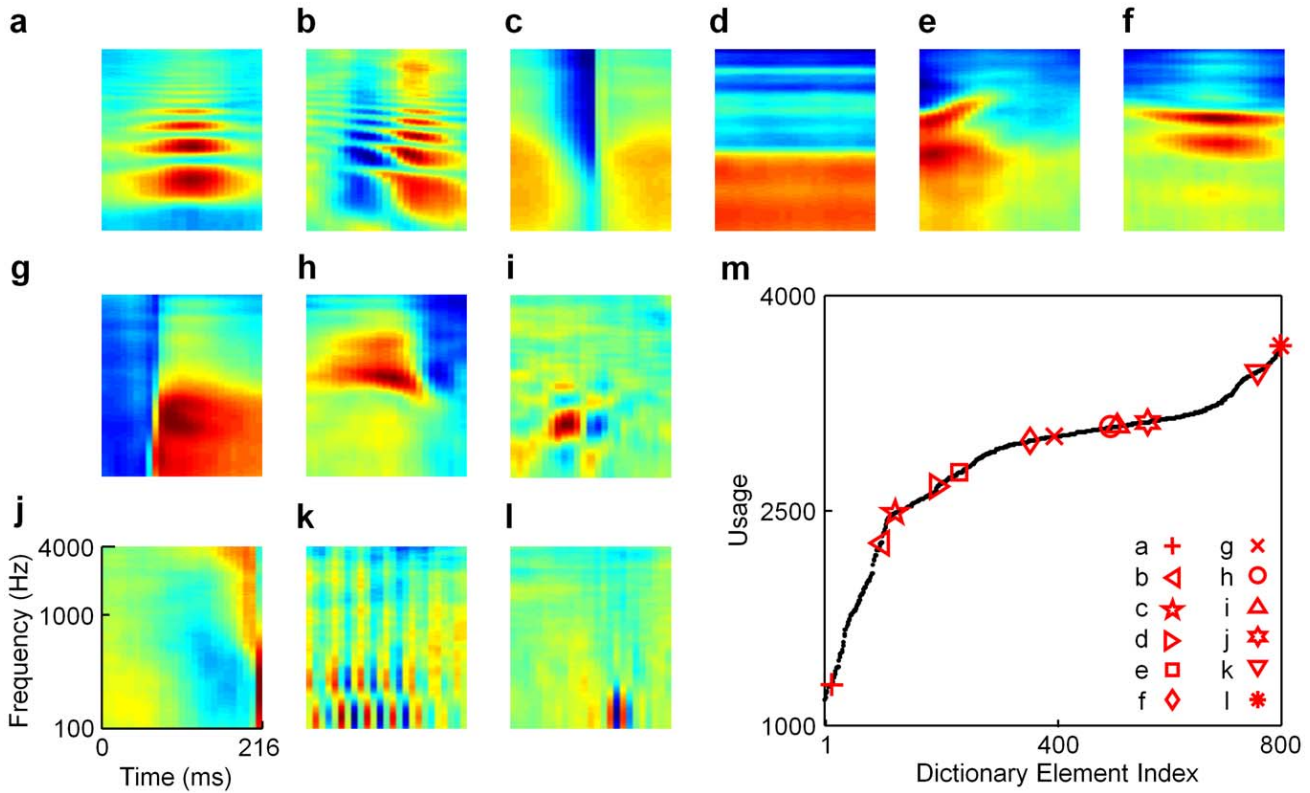

**Figure 4. A four-times overcomplete, L0-sparse dictionary trained on speech spectrograms.** This dictionary shows a greater diversity of shapes than the undercomplete dictionaries. (a–l) Representative elements **a**, **c**, **e**, **g**, **j**, and **l** resemble those of the half-complete dictionary (see **Fig. 3**). Other neurons display more complex shapes than those found in less overcomplete dictionaries: (**b**) a harmonic stack with flanking suppressive subregions; (**d**) a neuron sensitive to lower frequencies; (**f**) a short harmonic stack; (**h**) a localized but complex pattern of excitation with flanking suppression; (**i**) a localized checkerboard with larger excitatory and suppressive subregions than those in panel **l**; (**k**) a checkerboard pattern that extends for many cycles in time. Several of these patterns resemble neural spectro-temporal receptive fields (STRFs) reported in various stages of the auditory pathway that have not been predicted by previous theoretical models (see text and **Figs. 6–8**). (**m**) A graph of usage of the dictionary elements during inference. The different classes of dictionary elements still separate according to usage (see **Fig. S4** for the full dictionary) although the notable rises and plateaus as seen in **Fig. 3g** are less apparent in this larger dictionary.
doi:10.1371/journal.pcbi.1002594.g004

These same trends are present in the cochleogram-trained dictionaries. More types of STRFs appear when the degree of overcompleteness is increased (**Figs. S11, S12, S13**). For example, with more overcomplete dictionaries, some neurons have subfields spanning all frequencies or the full time-window within the cochleogram inputs. Additionally, we find neurons that exhibit both excitation and suppression in complex patterns, though the detailed shapes differ from what we find for the dictionaries trained on spectrograms.

We wondered to what extent the specific form of sparseness we imposed on the representation was affecting the particular features learned by our network. To study this, we used the LCA algorithm [24] to find the soft sparse solution (i.e., one that minimizes the L1 norm), and obtained similar results to what we found for the hard sparse cases: increasing overcompleteness resulted in greater diversity and complexity of learned features (see **Figs. S5, S6, S7, S8**). We also trained some networks using a different algorithm, called Sparsenet [10], for producing soft sparse dictionaries, and we again obtained similar results as for our hard sparse dictionaries (**Figs. S9, S10**). It has been proven mathematically [29] that signals that are actually L0-sparse can be uncovered effectively by L1-sparse coding algorithms, which suggests that speech is an L0-sparse signal given that we find similar features using algorithms designed to achieve either L1 or

L0 sparseness. Thus, preprocessing with spectrograms rather than a more nuanced cochlear model, and the degree of over-completeness, greatly influenced the learned dictionaries, unlike the different sparseness penalties we employed.

The specific form of the sparseness penalty did, however, affect the performance of the various dictionaries. In particular, the level of sparseness achieved across the population of model neurons exhibited different relationships with the fidelity of their representations, suggesting that some model choices resulted in population codes that were more efficient at using small numbers of neurons to represent stimuli efficiently, while others were more effective at increasing their representational power when incorporating more active neurons (**Fig. S20**).

## Modulation Power Spectra

Our four-times overcomplete, spectrogram-trained dictionary exhibits a clear tradeoff in spectrotemporal resolution (red points, **Fig. 5**), similar to what has been found experimentally in IC [30]. IC is the lowest stage in the ascending auditory pathway to exhibit such a complete tradeoff, but it has yet to be determined for higher stages of processing, such as A1. This trend is not present in the half-complete cochleogram-trained dictionary (blue open circles, **Fig. 5**). Rather, these elements display a limited range of temporal modulations, but they span nearly the full range of possible spectral modulations. Thus, by this measure the spectrogram-trained dictionary is a better model of IC than the cochleogram-trained model. In the next section, we compare the shapes of the various classes of model STRFs with individual neuronal STRFs from IC, and again find good agreement between our overcomplete spectrogram-trained model and the neural data.

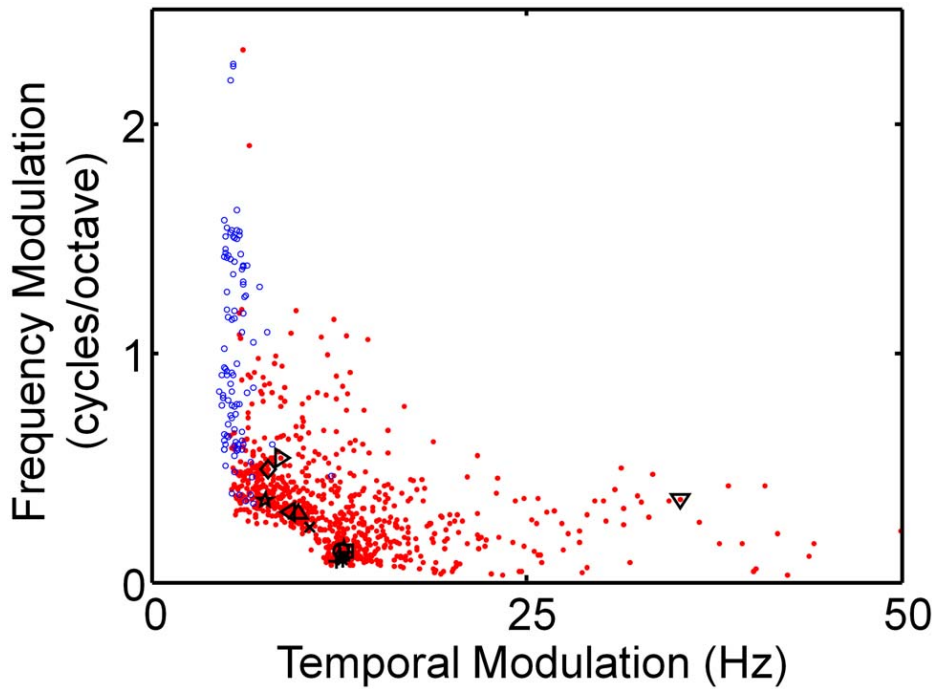

**Figure 5. Our overcomplete, spectrogram-trained model exhibits similar spectrotemporal tradeoff as Inferior Coliculus.** Modulation spectra of half-complete cochleogram-trained dictionary and four-times overcomplete spectroram-trained dictionary are shown. The four-times overcomplete spectrogram-trained dictionary elements (red dots; same dictionary as in **Fig. 4**) display a clear tradeoff between spectral and temporal modulations, similar to what has been reported for Inferior Colliculus (IC) [30]. By contrast, the half-complete cochleogram-trained dictionary (blue circles; same dictionary as in **Fig. 2**) exhibits a much more limited range of temporal modulations, with no such tradeoff in spectrotemporal resolution. Each data point represents the centroid of the modulation spectrum of the corresponding element. The elements shown in **Fig. 4** are indicated on the graph with the same symbols as before.
doi:10.1371/journal.pcbi.1002594.g005

### Predicting Acoustic Features that Drive Neurons in IC and Later Stages in the Ascending Auditory Pathway

Our model learns features that resemble STRFs reported in IC [30–33], as well as in the ventral side of the medial geniculate body (MGBv) [34] and A1 [34–36]. We are unaware of any previous theoretical work that has provided accurate predictions for receptive fields in these areas.

**Figs. 6**, **7**, and **8** present several examples of previously reported experimental receptive fields that qualitatively match some of our model's dictionary elements. We believe the most important class of STRFs we have found are localized checkerboard patterns of excitation and suppression, which qualitatively match receptive fields of neurons in IC and MGBv (**Fig. 7**).

IC neurons often exhibit highly localized excitation and suppression patterns (**Fig. 6**), sometimes referred to as "ON" or "OFF" responses, depending on the temporal order of excitation and suppression. We show multiple examples drawn from the complete, two-times overcomplete, and four-times overcomplete dictionaries, trained on spectrograms, that exhibit these patterns. The receptive fields of two neurons recorded in gerbil IC exhibit suppression at a particular frequency followed by excitation at the same frequency (**Fig. 6a**). Such neurons are found in our model dictionaries (**Fig. 6b**). The reverse pattern is also found in which suppression follows excitation as shown in two cat IC STRFs (**Fig. 6c**) with matching examples from our model dictionaries (**Fig. 6d**). Note that the experimental receptive fields extend to higher frequencies because the studies were done in cats and gerbils, which are sensitive to higher frequencies than we were probing with our human speech training set. The difference in time-scales between our spectrogram representation and the experimental STRFs could reflect the different timescales of speech and behaviorally relevant sounds for cats and rodents.

A common feature of thalamic and midbrain neural receptive fields is a localized checkerboard pattern of excitation and suppression (**Fig. 7**), typically containing between four to nine distinct subfields. We present experimental gerbil IC, cat IC and cat MGBv STRFs of this type in **Fig. 7a** beside similar examples from our model (**Fig. 7b**). This pattern is displayed by many elements in our sparse coding dictionaries, but to our knowledge it has not been predicted by previous theories.

We also find some less localized receptive fields that strongly resemble experimental data. Some model neurons (**Fig. 8b**) consist of a suppression/excitation pattern that extends across most frequencies, reminiscent of broadband OFF and ON responses as reported in cat IC and rat A1 (**Fig. 8a**).

Another shape seen in experimental STRFs of bat IC (top), and cat A1 (bottom; **Fig. 8c**) is a diagonal pattern of excitation flanked by suppression at the higher frequencies. This pattern of excitation flanked by suppression is present in our dictionaries (**Fig. 8d**), including at the highest frequencies probed. This type of STRF pattern is reminiscent of the two-dimentional Gabor-like patches seen in V1, which have been well captured by sparse coding models of natural scenes [10,11,26].

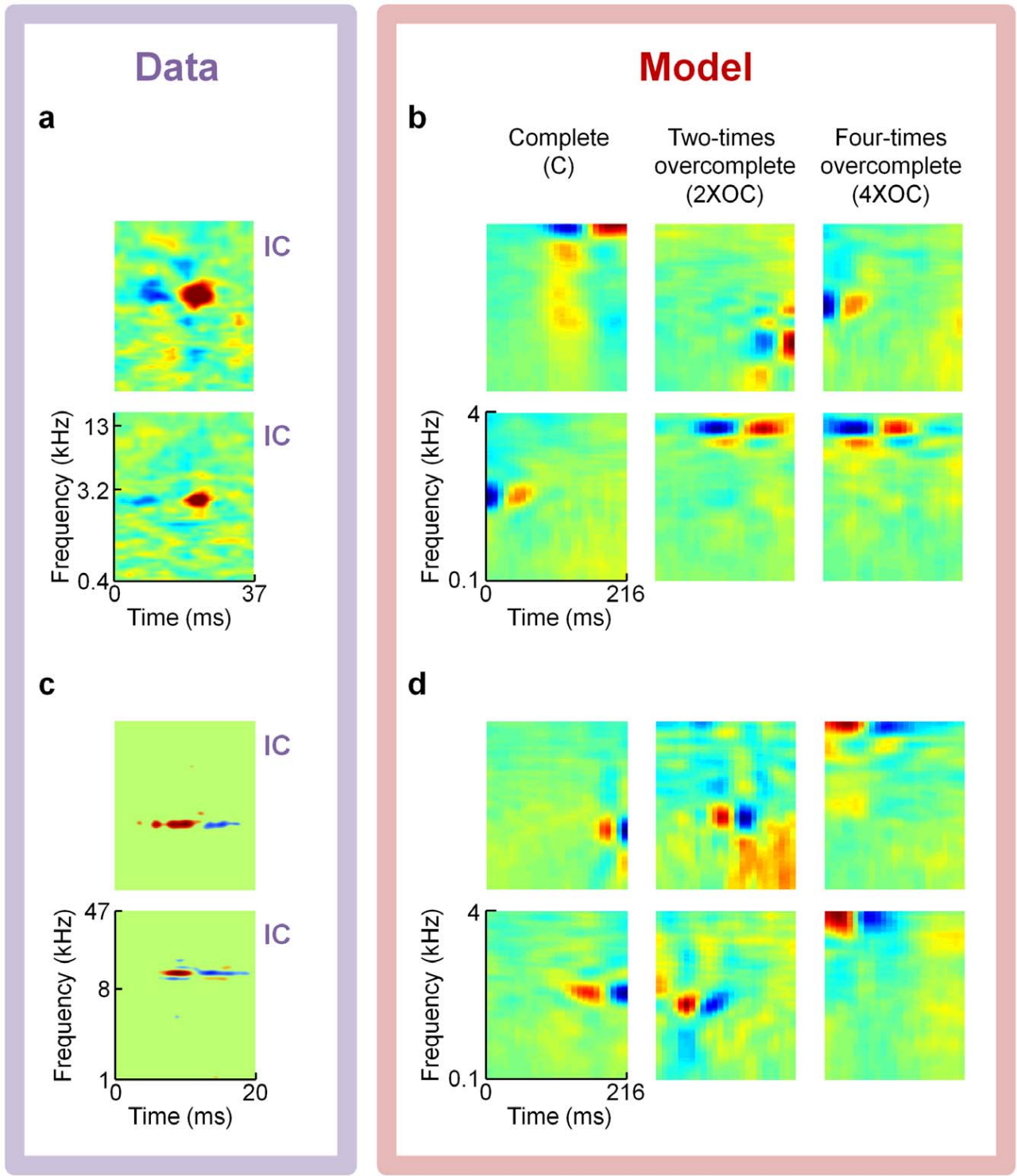

**Figure 6. Model comparisons to receptive fields from auditory midbrain.** Complete and overcomplete sparse coding models trained on spectrograms of speech predict Inferior Colliculus (IC) spectro-temporal receptive field (STRF) shapes with excitatory and suppressive subfields that are localized in frequency but separated in time. (**a**) Two examples of Gerbil IC neural STRFs [31] exhibiting ON-type response patterns with excitation following suppression; data courtesy of N.A. Lesica. (**b**) Representative model dictionary elements from each of three dictionaries that match this pattern of excitation and suppression. The three dictionaries were all trained on spectrogram representations of speech, using a hard sparseness (L0) penalty; the representations were complete (left column; **Fig. S2**), two-times overcomplete (middle column; **Fig. S3**), and four-times overcomplete (right column; **Fig. 4** and **Fig. S4**). (**c**) Two example neuronal STRFs from cat IC [30] exhibiting OFF-type patterns with excitation preceding suppression; data courtesy of M.A. Escabi. (**d**) Other model neurons from the same set of three dictionaries as in panel **b** also exhibit this OFF-type pattern.

doi:10.1371/journal.pcbi.1002594.g006

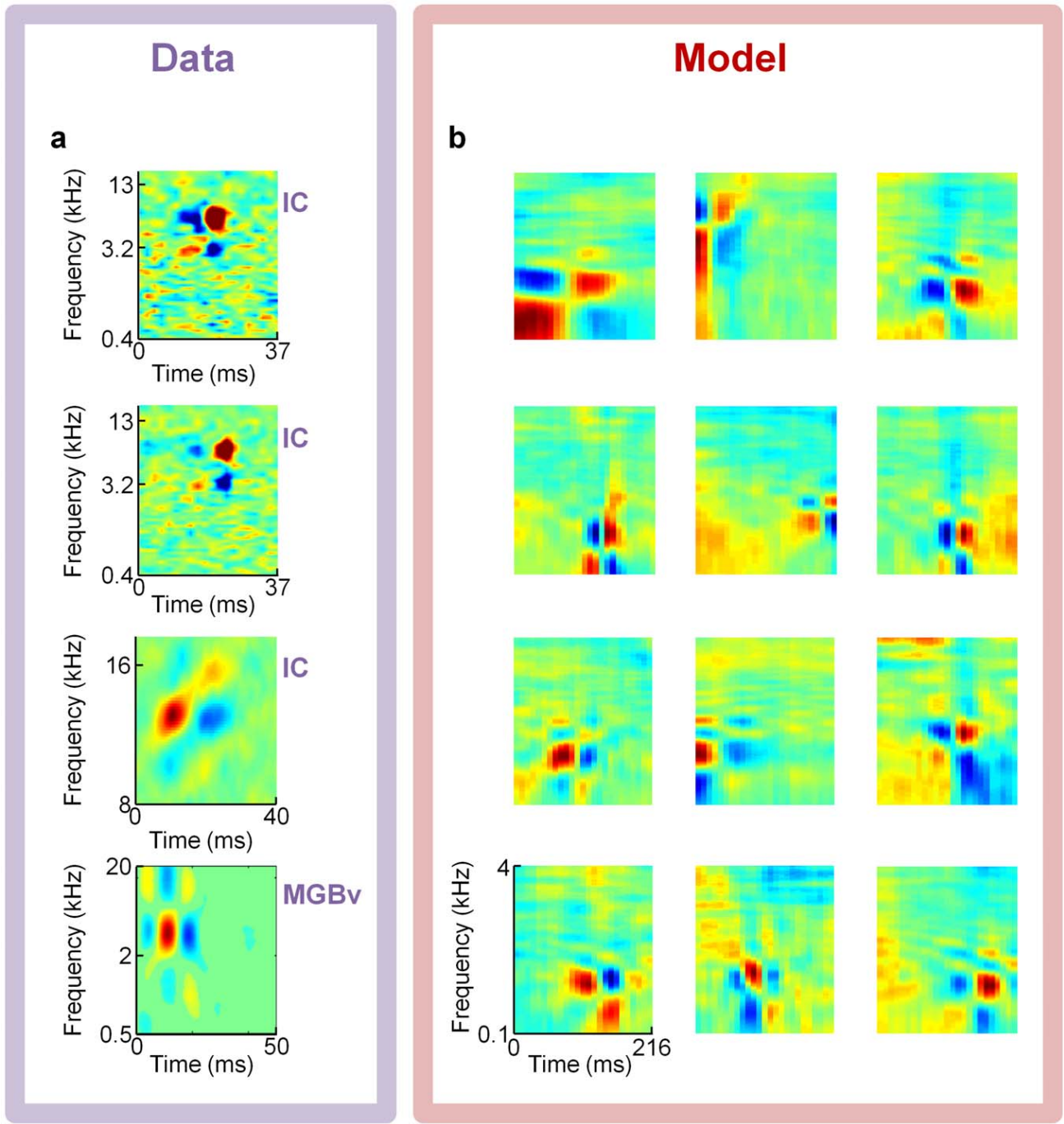

**Figure 7. Model comparisons to receptive fields from auditory midbrain and thalamus.** An overcomplete sparse coding model trained on spectrograms of speech predicts Inferior Colliculus (IC) and auditory thalamus (ventral division of the medial geniculate body; MGBv) spectro-temporal receptive fields (STRFs) consisting of localized checkerboard patterns containing roughly four to nine distinct subfields. (**a**) Example STRFs of localized checkerboard patterns from two Gerbil IC neurons [31], one cat IC neuron [33], and one cat MGBv neuron [34] (top to bottom). Data courtesy of N.A. Lesica (top two cells) and M.A. Escabí (bottom two cells). (**b**) Elements from the four-times overcomplete, L0-sparse, spectrogram-trained dictionary with similar checkerboard patterns as the neurons in panel **a**.
doi:10.1371/journal.pcbi.1002594.g007

## Discussion

We have applied the principle of sparse coding to spectrogram and cochleogram representations of human speech recordings in order to uncover some important features of natural sounds. Of the various models we considered, we have found that the specific form of preprocessing (*i.e.*, cochleograms vs. spectrograms) and the degree of overcompleteness are the most significant factors in determining the complexity and diversity of receptive field shapes. Importantly, we have also found that features learned by our sparse coding model resemble a diverse set of receptive field shapes in IC, as well as MGBv and A1. Even though a spectrogram may not provide as accurate a representation of the output of the cochlea as a more explicit cochleogram model, such as the one we explored here, we have found that sparse coding of spectrograms yields closer agreement to experimentally measured receptive fields, demonstrating that we can infer important aspects of sensory processing in the brain by identifying the statistically important features of natural sounds without having to impose many constraints from biology into our models from the outset.

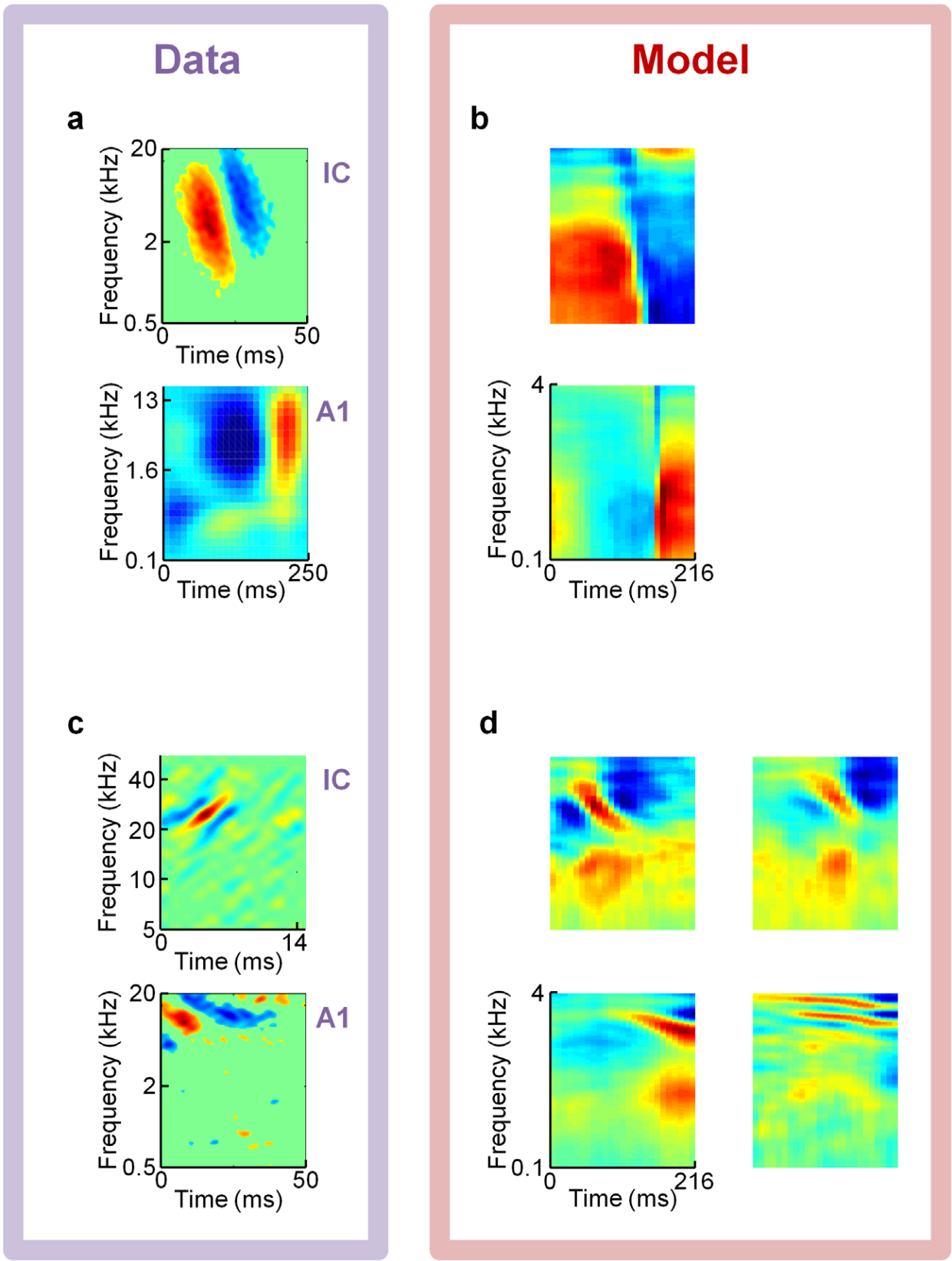

**Figure 8. Model comparisons to receptive fields from auditory midbrain and cortex.** on spectrograms of speech predicts several classes of broadband spectro-temporal receptive field (STRF) shapes found in Inferior Colliculus (IC) and primary auditory cortex (A1). (**a,b**) An example broadband OFF-type STRF from cat IC [34] (top; data courtesy of M.A. Escabí) and an example broadband ON-type subthreshold STRF from rat A1 [35] (bottom; data courtesy of M. Wehr) shown in panel **a** resemble example elements from a four-times overcomplete, L0-sparse, spectrogram-trained

Indeed, it is worth emphasizing that the agreement we have found did not result from fitting the neural physiology, *per se*; it emerged naturally from the statistics of the speech data we used to train our model. Specifically, the model parameters we explored — undercomplete vs. overcomplete representation, L0 vs. L1 sparseness penalty, and cochleogram vs. spectrogram preprocessing — represent a low-dimensional space of essentially eight different choices compared with the rich, high-dimensional space of potential STRF shapes we could have obtained.

Intriguingly, while we have emphasized the agreement between our model and IC, the receptive fields we have found resemble experimental data from multiple levels of the mammalian ascending auditory pathway. This may reflect the possibility that the auditory pathway is not strictly hierarchical, so that neurons in different anatomical locations may perform similar roles, and thus are represented by neurons from the same sparse coding dictionary. This view is consistent with the well-known observation that there is a great deal of feedback from higher to lower stages of processing in the sub-cortical auditory pathway [37], as compared with the visual pathway, for example. Some of our shapes have even been reported at lower levels. Harmonic stacks, including some with band-passed inhibition, have been reported in the dorsal cochlear nucleus [25,38] and they have been observed in presynaptic responses in IC (M.A. Escabí, C. Chen, and H. Read, Society for Neuroscience Abstracts 2011), but these shapes have not yet been reported in IC spiking responses or further up the ascending auditory pathway. The tradeoff in spectrotemporal resolution we have found in our model resembles that of IC, which is the lowest stage of the ascending auditory pathway to exhibit a tradeoff that cannot be accounted for by the uncertainty principle, as is the case for auditory nerve fibers [30], but it remains to be seen if such a tradeoff also exists in MGBv or A1.

A related issue is that an individual neuron might play different roles depending on the stimulus ensemble being presented to the nervous system. In fact, changing the contrast level of the acoustic stimuli used to probe individual IC neurons can affect the number of prominent subfields in the measured STRF of the neuron [31]. Our model does not specify which neuron should represent any given feature, it just predicts the STRFs that should be represented in the neural population in order to achieve a sparse encoding of the stimulus.

Moreover, for even moderate levels of overcompleteness, our sparse coding dictionaries include categories of features that have not been reported in the experimental literature. For example, the STRF shown in **Fig. 4k** represents a well-defined class of elements in our sparse dictionaries, but we are unaware of reports of this type of STRF in the auditory pathway. Thus, our theoretical receptive fields could be used to develop acoustic stimuli that might drive auditory neurons that do not respond to traditional probe stimuli. In particular, our dictionaries contain many broadband STRFs with complex structures. These broadband neurons may not have been found experimentally since by necessity researchers often probe neurons extensively with stimuli that are concentrated around the neuron's best frequency.

It is important to recognize that STRFs do not fully capture the response properties of neurons in IC, just as most of the explainable variance is not captured by linear receptive fields of V1 simple cells [39]. We note, however, that while our sparse coding framework involves a linear generative model, the encoding is non-linear. Thus, one of the questions addressed by this study is the degree to which the competitive nonlinearity of a highly over-complete model can account for the rich assortment of STRFs in IC. We have found that this is a crucial factor in learning a sparse representation that captures the rich variety of STRF shapes observed in IC, as well as in thalamus and cortex.

We have presented several classes of STRFs from our model that qualitatively match the shapes of neural receptive fields, but in many cases the neurons are sensitive to higher frequencies than the model neurons. This is likely due to the fact that we trained our network on human speech, which has its greatest energy in the low kHz range, whereas the example neural data available in the literature come from animals with hearing that extends to much higher acoustic frequencies, and with much higher-pitched vocalizations, than humans.

Our primary motivation for using speech came from the success of previous studies that yielded good qualitative [15] and quantitative [16] predictions of auditory nerve (AN) response properties based on sparse coding of speech. In fact, in order to obtain comparable results using environmental sounds and animal vocalizations, the relative proportion of training examples from each of three classes of natural sounds had to be adjusted to empirically match the results found using speech alone. Thus, speech provides a parameter-free stimulus set for matching AN properties, just as we have found for our model of IC. Moreover, good agreement between the model and AN physiology required selecting high SNR epochs within typically noisy recordings from the field; good results also required the selection of epochs containing isolated animal vocalizations rather than simultaneous calls from many individuals. By contrast, the speech databases used in those studies and the present study consist of clean, high SNR recordings of individual speakers. The issue of SNR is especially important for our study given that the dimensionality of our training examples is much higher (6,400 values for our spectrogram patches; 200 values after PCA) compared with typical vision studies (*e.g.*, 64 pixel values [10]).

Beyond the practical benefits of training on speech, the basic question of whether IC is best thought of as specialized for conspecific vocalizations or suited for more general auditory processing remains unanswered, but it seems reasonable to assume that it plays both roles. Questions such as this have inspired an important debate about the use of artificial and ecologically relevant stimuli [40,41] and what naturalistic stimuli can tell us about sensory coding [42–44]. The fact that several of the different STRFs we find have been observed in a variety of species, including rats, cats, and ferrets, suggests that there exist sufficiently universal features shared by the specific acoustic environments of these creatures to allow some understanding of IC function without having to narrowly tailor the stimulus set to each species.

Even if sparse coding is, indeed, a central organizing principle throughout the nervous system, it could still be that the sparse representations we predict with our model correspond best to the subthreshold, postsynaptic responses of the membrane potentials of neurons, rather than their spiking outputs. In fact, we show an example of a subthreshold STRF (**Fig. 8a** bottom) that agrees well with one class of broadband model STRFs (**Fig. 8b**). The tuning properties of postsynaptic responses are typically broader than spiking responses, as one would expect, which could offer a clue as to which is more naturally associated with model dictionary elements. If our model elements are to be interpreted as subthreshold responses, then the profoundly unresponsive regions surrounding the active subfields of the neuronal STRFs could be more accurately fit by our model STRFs after they are post-processed by being passed through a model of a spiking neuron with a finite spike threshold.

It is encouraging that sparse encoding of speech can identify acoustic features that resemble neuronal STRFs from auditory midbrain, as well as those in thalamus and cortex, and it is notable that the majority of these features bear little resemblance to the Gabor-like shapes and elongated edge detectors that have been predicted by sparse coding representations of natural images. Clearly, our results are not an unavoidable consequence of the sparse coding procedure itself, but instead reflect the structure of the speech spectrograms and cochleograms we have used to train our model. Previous work to categorize receptive fields in A1 has often focused on oriented features that are localized in time and frequency [27,45], and some authors have suggested that such Gabor-like features are the primary cell types in A1 [46], but the emerging picture of the panoply of STRF shapes in IC, MGBv, and A1 is much more complex, with several distinct classes of features, just as we have found with our model. An important next step will be to develop parameterized functional forms for the various classes of STRFs we have found, which can assume the role that Gabor wavelets have played in visual studies. We hope that this approach will continue to yield insights into sensory processing in the ascending auditory pathway.

## Methods

### Sparse Coding

In sparse coding, the input (spectrograms or cochleograms) $\mathbf{y}$ is encoded as a matrix $\mathbf{A}$ multiplied by a vector of weighting coefficients $\mathbf{s}$: $\mathbf{y} = \mathbf{A}\mathbf{s} + \varepsilon$ where $\varepsilon$ is the error. Each column of $\mathbf{A}$ represents one dictionary element or receptive field, the stimulus that most strongly drives the neuron. If there are more columns in $\mathbf{A}$ than elements in $\mathbf{y}$, this will be an overcomplete representation. We defined the degree of overcompleteness relative to the number of principle components. We learned the dictionary and inferred the coefficients by descending an energy function that minimizes the mean squared error of reconstruction under a sparsity constraint.

$$E(t) = \frac{1}{2} \|\mathbf{y}(t) - \mathbf{A}\mathbf{s}(t)\|^2 + \lambda \sum_m C(s_m(t)). \qquad (1)$$

Here $\lambda$ controls the relative weighting of the two terms and $C$ represents the sparsity constraint.

The sparsity constraint requires the column vector $\mathbf{s}$ to be sparse by some definition. In this paper, we focus on the L0-norm, minimizing the number of non-zero coefficients in $\mathbf{s}$ (or equivalently the number of active neurons in a network). Another norm we have investigated is the L1-norm, minimizing the absolute activity of all of the neurons.

### Locally Competitive Algorithm

We performed inference of the coefficients with a recently developed algorithm, a Locally Competitive Algorithm [24], which minimizes close approximations of either the L0- or L1-norms. Each basis function $\mathbf{A}_i$ is correlated with a computing unit defined by an internal variable $u_i$ as well as the output coefficient $s_i$. All of the neurons begin with the coefficients set to zero. These values change over time depending on the input. A neuron $u_i$ increases by an amount $b_i$ if the input overlaps with the receptive field of the neuron: $b_i(t) = \langle \mathbf{A}_i, \mathbf{y}(t) \rangle$. The neurons evolve as a group following dynamics in which the neurons compete with one another to represent the input. The neurons inhibit each other with the strength of the inhibition increasing as the overlap of their receptive fields and the output coefficient values increase. This internal variable is then put through a thresholding function $T_\lambda$ to produce the output value: $s_i = T_\lambda(u_i)$.

In vector notation, the full dynamic equation of inference is:

$$\dot{\mathbf{u}}(t) = f(\mathbf{u}(t)) = \frac{1}{\tau} [\mathbf{b}(t) - \mathbf{u}(t) - (\mathbf{A}^T \mathbf{A} - I)\mathbf{s}(t)],$$
$$\mathbf{s}(t) = T_\lambda(\mathbf{u}(t)). \qquad (2)$$

The variable $\tau$ sets the time-scale of the dynamics.

The thresholding function $T_\lambda$ is determined by the sparsity constraint $C$. It is specified via:

$$\lambda \frac{dC(s_m)}{ds_m} = u_m - s_m = u_m - T_\lambda(u_m). \qquad (3)$$

### Learning

Learning is done via gradient descent on the energy function:

$$\mathbf{r}(t) = \mathbf{y}(t) - \mathbf{A}\mathbf{s}(t),$$
$$\mathbf{A} = \mathbf{A} + \eta_{\mathbf{A}}(\mathbf{r}(t)\mathbf{s}^T(t)) + \theta(\mathbf{A} - \mathbf{A}\mathbf{A}^T\mathbf{A}). \qquad (4)$$

The $\theta$ term is a device for increasing orthogonality between basis functions [47]. This is equivalent to adding in a prior that the basis functions are unique.

### Stimuli

We used two corpora of speech recordings from the handbook of the International Phonetic Association (http://web.uvic.ca/ling/resources/ipa/handbook_downloads.htm) and TIMIT [48]. These consist of people telling narratives in approximately 30 different languages. We resampled all waveforms to 16000 Hz, and then converted them into spectrograms by taking the squared Fourier Transform of the raw waveforms. We sampled at 256 frequencies logarithmically spaced between 100 and 4000 Hz. We monotonically transformed the output with the logarithm function, resulting in the log-power of the sound at specified frequencies over time.

The data was then divided into segments covering all frequencies and 25 overlapping time points (16 ms each) representing 216 ms total. Subsequently, we performed principal components analysis on the samples to whiten the data as well as reduce the dimensionality. We retained the first 200 principal components as this captured over 93% of the variance in the spectrograms and lowered the simulation time. During analysis, the dictionaries were dewhitened back into spectrogram space.

We also trained with another type of input, cochleograms [21,22]. These are similar to spectrograms, but the frequency filters mimic known properties of the cochlea via a cochlear model [21]. The cochlear model sampled at 86 frequencies between 73 and 7630 Hz. For this input, the total time for each sample was 250 ms (still 25 time points), and the first 200 principle components captured over 98% of the variance.

## Presentation of Dictionaries

All dictionary neurons were scaled to be between −1 and 1 when displayed. The coefficients in the encoding can take on positive or negative values during encoding. To reflect this, we looked at the skewness of each dictionary element. If the skewness was negative, the colors of the dictionary element were inverted when being displayed to reflect the way that element was actually being used.

## Modulation Power Spectra

To calculate the modulation power spectra, we took a 2D Fourier Transform of each basis function. For each element, we plotted the peak of the temporal and spectral modulation transfer functions (**Fig. 5**). For the cochleogram-trained basis functions, we approximated the cochleogram frequency spacing as being log-spaced to allow comparison with the spectrogram-trained dictionaries.

## Presentation of Experimental Data

Data from [31] was given to us in raw STRF format. Each was interpolated by a factor of three, but no noise was removed. Data from [30,32–35] were given to us in the same format as they were originally published.

## Supporting Information

**Figure S1 The full set of elements from a half-complete, L0-sparse dictionary trained with LCA [24] on spectrograms of speech.** Each rectangle represents the spectrotemporal receptive field of a single element in the dictionary; time is plotted along the horizontal axis (from 0 to 216 ms), and log frequency is plotted along the vertical axis, with frequencies ranging from 100 Hz to 4000 Hz. Color indicates the amount of power present at each frequency at each moment in time, with warm colors representing high power and cool colors representing low power. Each element has been normalized to have unit Euclidean length. Elements are arranged in order of their usage during inference with usage increasing from left to right along each row, and all elements of lower rows used more than those of higher rows.
(TIFF)

**Figure S2 The full set of elements from a complete, L0-sparse dictionary trained with LCA [24] on spectrograms of speech.** Same conventions as Fig. S1.
(TIF)

**Figure S3 The full set of elements from a two times overcomplete, L0-sparse dictionary trained with LCA [24] on spectrograms of speech.** Same conventions as Fig. S1.
(TIF)

**Figure S4 The full set of elements from a four times overcomplete, L0-sparse dictionary trained with LCA [24] on spectrograms of speech.** Same conventions as Fig. S1.
(TIF)

**Figure S5 The full set of elements from a half-complete, L1-sparse dictionary trained with LCA [24] on spectrograms of speech.** Same conventions as Fig. S1.
(TIF)

**Figures S6 The full set of elements from a complete, L1-sparse dictionary trained with LCA [24] on spectrograms of speech.** Same conventions as Fig. S1.
(TIF)

**Figure S7 The full set of elements from a two times overcomplete, L1-sparse dictionary trained with LCA [24] on spectrograms of speech.** Same conventions as Fig. S1.
(TIF)

**Figure S8 The full set of elements from a four times overcomplete, L1-sparse dictionary trained with LCA [24] on spectrograms of speech.** Same conventions as Fig. S1.
(TIF)

**Figure S9 The full set of elements from a half-complete, L1-sparse dictionary trained with Sparsenet [10] on spectrograms of speech.** Same conventions as Fig. S1.
(TIF)

**Figure S10 The full set of elements from a complete, L1-sparse dictionary trained with Sparsenet [10] on spectrograms of speech.** Same conventions as Fig. S1.
(TIF)

**Figure S11 The full set of elements from a complete, L0-sparse dictionary trained using LCA [24] on cochleograms of speech.** Each rectangle represents the spectrotemporal receptive field of a single element in the dictionary; time is plotted along the horizontal axis (from 0 to 250 ms), and log frequency is plotted along the vertical axis, with frequencies ranging from 73 Hz to 7630 Hz. Color indicates the amount of power present at each frequency at each moment in time, with warm colors representing high power and cool colors representing low power. Each element has been normalized to have unit Euclidean length. Elements are arranged in order of their usage during inference with usage increasing from left to right along each row, and all elements of lower rows used more than those of higher rows.
(TIF)

**Figure S12 The full set of elements from a two times overcomplete, L0-sparse dictionary trained with LCA [24] on cochleograms of speech.** Same conventions as Fig. S11.
(TIF)

**Figure S13 The full set of elements from a four times overcomplete, L0-sparse dictionary trained with LCA [24] on cochleograms of speech.** Same conventions as Fig. S11.
(TIF)

**Figure S14 The full set of elements from a half-complete, L1-sparse dictionary trained with LCA [24] on cochleograms of speech.** Same conventions as Fig. S11.
(TIF)

**Figure S15 The full set of elements from a complete, L1-sparse dictionary trained with LCA [24] on cochleograms of speech.** Same conventions as Fig. S11.
(TIF)

**Figure S16 The full set of elements from a two times overcomplete, L1-sparse dictionary trained with LCA [24] on cochleograms of speech.** Same conventions as Fig. S11.
(TIF)

**Figure S17 The full set of elements from a four times overcomplete, L1-sparse dictionary trained with LCA [24] on cochleograms of speech.** Same conventions as Fig. S11.
(TIF)

**Figure S18 The full set of elements from a half-complete, L1-sparse dictionary trained with Sparsenet [10] on cochleograms of speech.** Same conventions as Fig. S11.
(TIF)

**Figure S19 The full set of elements from a complete, L1-sparse dictionary trained with Sparsenet [10] on cochleograms of speech.** Same conventions as Fig. S11.
(TIF)

**Figure S20 The signal to noise ratio (SNR) of sparse coding dictionaries increases with overcompleteness and with increasing numbers of active elements.** Blue lines with triangles represent L0-sparse dictionaries, whereas green lines represent L1-sparse dictionaries. As expected, representations are more accurate with increasing numbers of active neurons and also when the level of overcompleteness is increased. Interestingly, the L0-sparse dictionaries typically have higher SNRs than the L1-sparse dictionaries. A few other general trends are evident as well.

Most notably, the L0-sparse dictionaries have higher SNRs than the L1-sparse dictionaries for similar levels of sparseness. Also, the more overcomplete dictionaries have higher SNRs than half-complete ones, even with the same absolute number of active neurons. The half-complete and complete dictionaries do not show much improvement in performance even as the number of active neurons increases. Interestingly, we find that the performance of the L0-sparse dictionaries tend to saturate as the fraction of active neurons approaches unity whereas the corresponding curves for the L1-sparse dictionaries tend to curve upwards. Note that we did not optimize the dictionaries at each data point, but instead used the same parameters used when training the network.
(TIF)

## Acknowledgments

The authors would like to thank J. Wang and E. Bumbacher for providing computer code and for helpful discussions. We are grateful to the following authors for sharing their neural data: F.A. Rodríguez, H.L. Read, M.A. Escabí, S. Andoni, N. Li, G.D. Pollak, N.A. Lesica, B. Grohe, A. Qiu, C.E. Schreiner, C. Machens, M. Wehr, H. Asari, and A.M. Zador. We thank M.A. Escabí, H.L. Read, A.M. Zador, and all members of the Redwood Center for useful discussions, and we thank K. Körding for helpful discussions and comments on the manuscript.

## Author Contributions

Conceived and designed the experiments: NLC MRD. Performed the experiments: NLC. Analyzed the data: NLC VLM MRD. Wrote the paper: NLC MRD.